\documentclass[aps,prl,twocolumn,groupedaddress,showpacs]{revtex4}

\usepackage{graphicx}
\usepackage{dcolumn}
\usepackage{bm}

\begin {document}
\title {Efficient rapid production of a Bose-Einstein condensate by overcoming serious three-body loss}

\author {Tetsuya Mukai}

\author {Makoto Yamashita}

\affiliation {NTT Basic Research Laboratories, NTT Corporation, 3-1, Morinosato-Wakamiya, 
Atsugi-shi, Kanagawa 243-0198, Japan}

\date {\today}

\begin {abstract}
We report the efficient production of a large Bose-Einstein condensate in $^{87}$Rb atoms. This is achieved by quickly reducing the radio-frequency of the magnetic field at a rate of $-96.8$~kHz/s during the final stage of evaporative cooling, and we have produced a condensate with ${(2.2 \pm 0.1) \times 10^{6}}$ atoms against a serious three-body recombination loss. We observed the dependence of the cooling efficiency on the rate at which the truncated energy changes by measuring the condensate growth for three kinds of radio-frequency sweep. The experimental results quantitatively agree with calculations based on the quantum kinetic theory of a Bose gas.
\end {abstract}

\maketitle
\section {Introduction}
Evaporative cooling is an essential experimental technique with regard to the recently demonstrated Bose-Einstein condensate (BEC) in magnetically trapped atomic gases \cite{Rb, Li, Na, H, He}. This cooling method is based on both the selective removal of energetic atoms through evaporation and collisional rethermalizations among the remaining atoms. However, the method is also subject to undesirable losses, e.g., losses caused by elastic collisions with the background gas, inelastic collisions due to dipolar relaxation, and three-body recombination processes \cite{eva, Luit, wal}. For alkali-metal atoms, such as $^{87}$Rb and $^{23}$Na, three-body recombination becomes a dominant loss mechanism during the final stage of evaporative cooling \cite{eva}, and limits the number of condensed atoms $N_{0}$. It is known that $^{87}$Rb atoms suffer much severer losses than $^{23}$Na atoms due to their large three-body recombination loss rate \cite{Rb_3body,Na_3body}. BEC experiments have clearly reflected this fact: the reported $N_{0}$ value in $^{87}$Rb is about $10^5$ on average and $10^6$ at most \cite{BEC_WWW, Rbmax, Rb_MIT}, while the relevant $N_{0}$ in $^{23}$Na is one order of magnitude larger than that ($10^6\sim10^7$) \cite{Na, Namax}. The small BEC of $^{87}$Rb is a great disadvantage not only for the precise measurement of condensate properties but also for the practical application of BEC.

The successful utilization of interatomic elastic collisions against loss mechanisms makes evaporative cooling highly efficient, and we can control the cooling efficiency by suitably changing the truncated energy $\epsilon_{t}$. With the forced evaporative cooling of atoms in a magnetic trap, resonant radio-frequency $\nu_{r\!f}$ is a good parameter with which to control the truncated energy through the equation $\epsilon_{t}=|m_{F}|h(\nu_{r\!f}-\nu_{0})$, where $m_{F}$ is the magnetic quantum number, $h$ is the Planck constant and $\nu_{0}$ is the frequency corresponding to the trap minimum \cite{eva, Sackett, 1D_ev, Simplified_System}. In order to overcome the serious three-body loss that occurs in evaporative cooling and to achieve a large $N_{0}$, we have calculated the optimized radio-frequency (RF) sweep based on the quantum kinetic theory of a Bose gas \cite{Opt_Yama, Yamashita_2000, Yamashita_1999}, which can provide the quantitative results on the whole evaporative cooling process of the experiments. According to this calculation, the fast RF sweep around the BEC transition point provides highly efficient evaporative cooling. The optimized RF sweep is a complicated function of many experimental parameters that are specific to each experimental setup, e.g., the temperature of atomic cloud, the density of atoms, the atomic collisional parameters, the magnetic trapping confinement, and so on. As a general guideline for practical optimization in experiments, it was also demonstrated that a fast linear RF sweep during the final stage of the evaporative cooling process is very effective for realizing a large $N_{0}$ in $^{87}$Rb atoms \cite{Opt_Yama}.

In this paper we experimentally applied a fast linear RF sweep during the final stage of evaporative cooling to $^{87}$Rb atoms in the ${|F=2, m_{F}=+2~\rangle}$ state and successfully demonstrated the efficient rapid growth of a large number of condensed atoms against serious three-body loss. The experimental result quantitatively agrees with our calculations based on the quantum kinetic theory of a Bose gas. The three-body recombination loss, which has been observed in the decay rate of a condensate \cite{Rb_3body, Na_3body}, is experimentally studied by observing the dependence of the growth of a condensate \cite{Growth_Na, Growth_Rb} on the rate at which the truncated energy changes.

\section {Experimental apparatus}
Our experimental setup is based on the double magneto-optical trap (DMOT) configuration: a chamber for the first MOT, located in the top half of the setup, is connected to a glass cell in the bottom half for the second MOT with a narrow tube (inside diameter 10~mm, length 343~mm, inclined ${30^\circ}$ from the horizontal plane) surrounded by a multi-pole magnetic field. The small conductance of the narrow tube enables us to realize a sufficiently low background gas pressure, ${(2.2 \pm 0.2) \times 10^{-11}}$~torr, in the lower glass cell while the pressure of the upper chamber is high enough to produce an atomic cloud every 67~ms in the first MOT. The light for the experiment is provided by Coherent 899 Ti:S lasers at $^{87}$Rb resonance (780~nm). The experimental sequence begins by loading $^{87}$Rb atoms in the first MOT. We then push the trapped atoms toward the lower glass cell with 15~Hz resonant pulses. In the lower glass cell, atoms are collected in the second MOT for 30~s, cooled to about 30~${\mu}$K by polarization gradient cooling, and optically pumped back to ${|F=2, m_{F}=+2~\rangle}$ state. After these processes, we turn off all the laser lights and turn on a shallow magnetic potential in 0.3~ms. We then ramp up the magnetic potential adiabatically to its maximum confinement in 3.5~s. At this stage we collect $N=5 \times 10^{8}$ atoms in a magnetic trap at temperature $T=500~{\mu}$K.

The magnetic trap is a cloverleaf Ioffe-Pritchard type made of a narrow copper tube (2~mm in diameter and 0.5~mm thick). The axial curvature $B_{z}''$ and the radial gradient $B_{\rho}'$ of the magnetic field at the trap center are $B_{z}''=172$~G/cm$^{2}$ and $B_{\rho}'=132$~G/cm at a driving current of $230$~A.  By atom laser output coupling \cite{AtomLaser}, we observe $4.6$~kHz/s upward drift of the trap minimum after driving with a large current for $45$~s, and the bias of the magnetic field is $B_{0}=(1.36\sim1.39)$~G. Periodic operations of the whole experimental sequence and a high power chilling system with temperature controlled water enable us to stabilize the shot-to-shot fluctuations of the upward drift of the trap minimum to less than ${\pm}0.2$~kHz. The influence of the upward drift of the trap minimum on the oscillation frequency of the trap is less than $0.5$~\% and the axial oscillation frequency is $\omega_{z}=2\pi \times 16.7$~Hz and the radial oscillation frequency  is $\omega_{\rho}=2\pi \times 142$~Hz.

We obtain a time-of-flight density profile of atoms after $18$~ms of ballistic expansion by injecting an expanded resonant pulse ($0.2$~mW/cm${^{2}}$, $50~\mu$s) in the vertical direction. $N_{0}$ is extracted from the absorption profile fitted by the Thomas-Fermi approximation and it is cross-checked with the $N_{0}$ value calculated from the magnetic confinement and the chemical potential \cite{BEC_measurement}.

%
\begin{figure}
\begin{center}
\includegraphics[height=5.5cm]{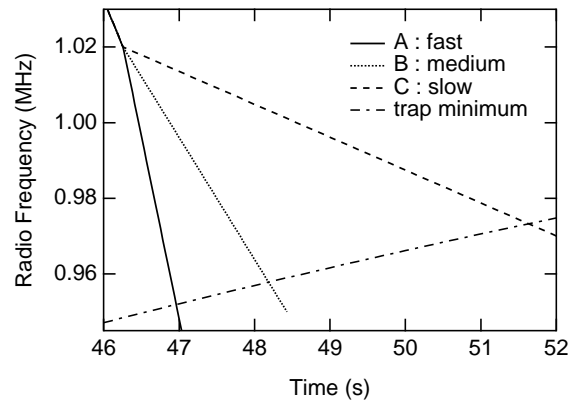}
\caption{\label{Sweep} Time dependence of radio-frequency during the final stage of evaporative cooling used in this experiment. The origin of the time scale represents the starting point of the evaporative cooling. Frequency changing rate for each RF sweep is $-96.8$~kHz/s (A: fast), $-32.3$~kHz/s (B: medium), and $-8.7$~kHz/s (C: slow), respectively. The drift of the trap minimum is $4.6$~kHz/s.}
\end{center}
\end{figure}

After the recapture and compression of atoms in the magnetic trap we start evaporative cooling with applying an RF magnetic field. The RF is controlled from $\nu_{r\!f}=36$~MHz to $1.02$~MHz in $46.258$~s according to the optimized calculation based on the quantum kinetic theory of a Bose gas \cite{Opt_Yama}.  At $\nu_{r\!f}=1.02$~MHz, we collect $N=(7.7\pm0.4)\times10^{6}$ atoms at temperature $T=(951\pm50)~\mu$K, i.e., the phase space density is evaluated to be $\rho_{ps}={0.57\pm0.10}$ and the system is set very close to the BEC transition point.

At a high phase space density the three-body recombination becomes the dominant loss mechanism of evaporative cooling. It is known that the time constant of the loss rate caused by three-body recombination is of the order of a second, and the fast linear RF sweep during the final stage of the evaporative cooling process is predicted to be very effective in increasing $N_{0}$ \cite{Opt_Yama}. In order to demonstrate efficient evaporative cooling and obtain a large $N_{0}$, we employed three kinds of linear RF sweep with different frequency changing rates as $-96.8$~kHz/s (A: fast), $-32.3$~kHz/s (B: medium), and $-8.7$~kHz/s (C: slow) during the final $71$~kHz (Fig.~\ref{Sweep}). Including the drift of the trap minimum $4.6$~kHz/s, the truncated energy changing rate $\dot{\epsilon_{t}}$ for each RF sweep is $-9.7~\mu$K/s (A: fast), $-3.5~\mu$K/s (B: medium), and $-1.3~\mu$K/s (C: slow), respectively. The forced evaporative cooling is well scaled by the characteristic time $\tau$ given by the inverse of the normalized truncated energy changing rate ($\tau=\left|\dot{\epsilon_{t}}/{\epsilon_{t}}\right|^{-1}$). The following inequality
\begin{equation}
\label{Quasi_Static}
\tau\gg{2\pi}/{\omega_{z}},~{2\pi}/{\omega_{\rho}},~\tau_{coll},
\end{equation}
where $\tau_{coll}$ ($\le$ 10~ms) is the elastic collision time, is a criterion of quasi-static condition, and the violation of this inequality will introduce the nonequilibrium effects into the evaporative cooling process as reported in Ref. \cite{None_Equilib}. We have selected the smallest $\tau$ in this experiment, i.e., $\tau=0.71$~s for the fastest sweep, to be one order larger than the criterion of Eq.~(\ref{Quasi_Static}). For this reason all the forced evaporative cooling in this experiment is quasi-static and the quick rethermalization condition is satisfied. On the other hand, with quasi-static evaporative cooling, the truncation parameter $\eta = \epsilon_{t}/k_BT$, where $k_{B}$ is the Boltzmann constant and $T$ is the system temperature, changes very slowly during the cooling process \cite{eva, wal}. By observing the temperature variation during the evaporative cooling process, the truncation parameters for the fast, medium, and slow RF sweeps are measured and found to be almost constant at $\eta=6.0$, $\eta=6.7$, and $\eta=7.8$, respectively. These measurements also confirm the quasi-static property of the present evaporative cooling.

\section{Results and discussion}
%
%
\begin{figure}
\begin{center}
\includegraphics[height=5.5cm]{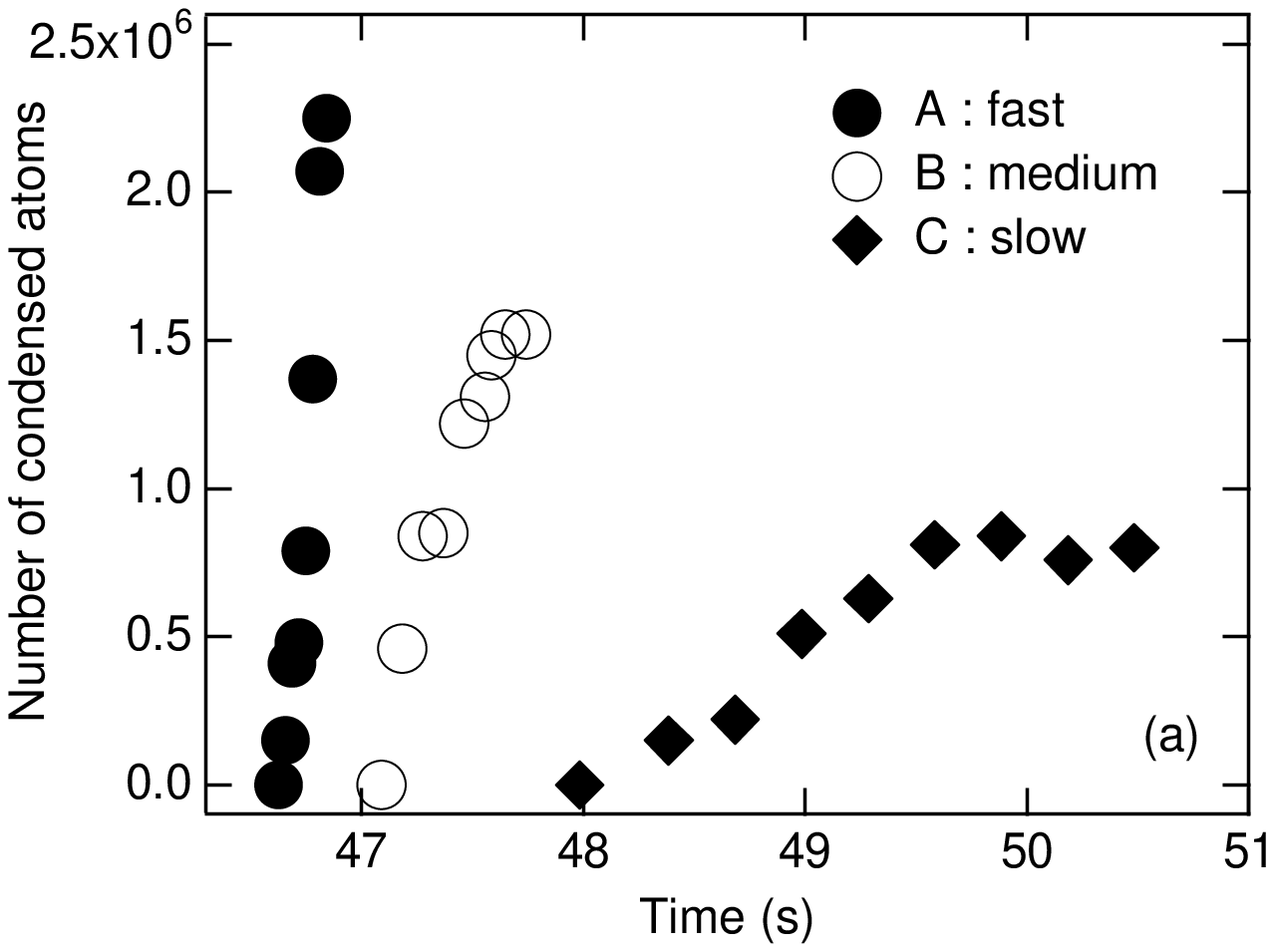}
\includegraphics[height=5.5cm]{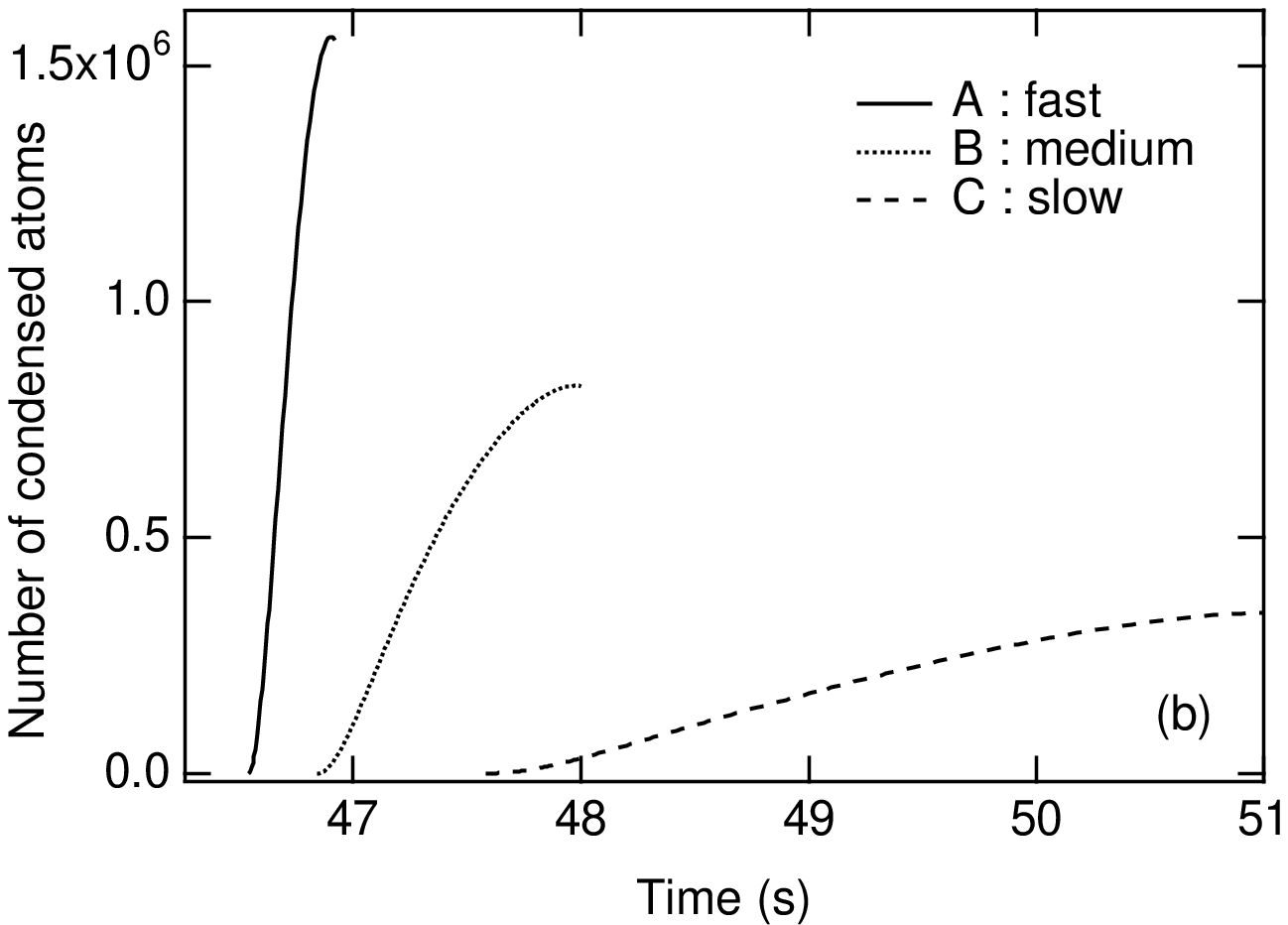}
\caption{\label{Growth_T} The increase in the number of condensed atoms $N_{0}$ for the three kinds of RF sweep; (a) experiment and (b) calculation. The experimental (calculated) data of the fast, medium, and slow RF sweeps are represented by the filled circles (solid line), open circles (dotted line), and filled diamonds (dashed line), respectively. The decay rate constants due to the background gas collisions, dipolar relaxation, and three-body recombination adopted in this calculation are $G_{1}$=1/70~s${^{-1}}$, $G_{2}$=1.0${\times}$10${^{-15}}$~cm${^{3}}$/s \cite{dipolar}, and $G_{3}$=9.0${\times}$10${^{-29}}$~cm${^{6}/s}$ \cite{Rb_3body}, respectively.}
\end{center}
\end{figure}

Figure~\ref{Growth_T}(a) is the experimentally observed increase in $N_{0}$ for the three kinds of RF sweep. With a fast RF sweep (filled circles), the atoms start to condense at $46.625$~s and $N_{0}$ rapidly increases to its maximum value, $(2.2\pm0.1)\times10^{6}$, in 0.217~s. The maximum $N_{0}$ for the fast RF sweep is much larger than those for the other two RF sweeps, $(1.5\pm0.1)\times10^{6}$ and $(8.0 \pm 0.6) \times 10^{5}$, and the condensate growth process finishes in a very short time. With the medium (slow) RF sweep, shown by the open circles (filled diamonds), it takes 0.558~s (1.612~s) for $N_{0}$ to reach its maximum value and $N_{0}$ seems to be saturated after 47.648~s (49.6~s). Figure~\ref{Growth_T}(b) shows the results of calculations based on the quantum kinetic theory of a Bose gas using the experimental parameters. Both the order of $N_{0}$ and the transition time of the experimental data agree quantitatively with our calculation without using any fitting parameters. Discrepancy between the experimental and the theoretical $N_{0}$ values might be due to some ambiguities of the collisional parameters used in the calculations ($G_{1}$, $G_{2}$, and $G_{3}$), the simplification of our cloverleaf magnetic trap into the Ioffe-Pritchard type, and the negligence of the gravitational sag of the magnetic trap \cite{eva, wal}.

Figure 2 proves that the fast linear RF sweep during the final stage of evaporative cooling works well in obtaining a large $N_{0}$ in $^{87}$Rb atoms as we have predicted theoretically in Ref. \cite{Opt_Yama}. This result is consistent with the recent report on the full experimental optimization of evaporative cooling in Ref. \cite{Simplified_System}. Furthermore, our present experiment clearly points out the importance of the final stage of evaporative cooling. Thus we can expect, with a much slower RF sweep such as the commonly used exponential function, the evaporative cooling will become inefficient and the maximum $N_{0}$ will be limited to a smaller value \cite{Stab_Yama}.

%
\begin{figure}
\begin{center}
\includegraphics[height=6cm]{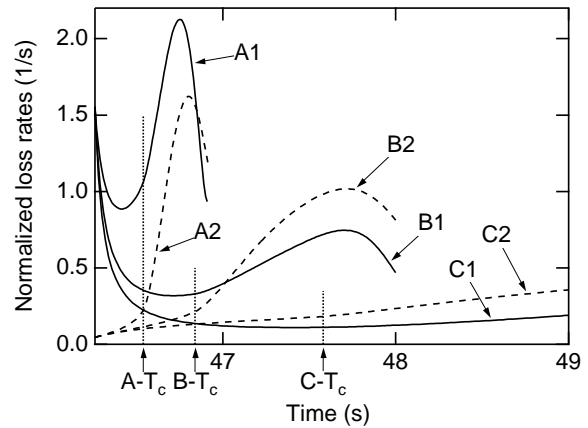}
\caption{\label{Loss_rate} Loss rates for the three kinds of RF sweep normalized by the total number of trapped atoms at each time. A, B, and C represent the fast, medium, and slow RF sweep, respectively. The solid curves (A1, B1, and C1) represent the loss rates caused by evaporation and the dashed ones  (A2, B2, and C2) represent those caused by three-body recombination, respectively. A-T${_{c}}$, B-T${_{c}}$, and C-T${_{c}}$ represent the BEC transition point for each RF sweep.}
\end{center}
\end{figure}

Next we discuss the dynamics of evaporative cooling and try to understand the experimental results with the loss rate plot in Fig.~\ref{Loss_rate}, which is a plot of the normalized loss rate caused by evaporation (solid lines) and three-body recombination (dashed lines) for the three kinds of RF sweep (A:~fast, B:~medium, and C:~slow). The loss rate curves for the fast RF sweep (A1 and A2) represent typical loss rate behavior and we will use them to understand the dynamics of evaporative cooling. Loss rates caused by evaporation and three-body recombination start both to increase slightly before the BEC transition (A-T$_{c}$) because of the high density of atoms. After reaching the critical point, the fast growth of the condensate by bosonic stimulation introduces a sudden increase in the local density of atoms and both loss rates rise abruptly.  After sufficient condensate growth, the number of atoms in the thermal fraction is small and the loss rates start to decrease. This is because only atoms in a thermal fraction can evaporate, and the three-body decay rate of atoms in a condensate is $3!=6$ times smaller than that of atoms in a thermal fraction \cite{kagan, Rb_3body}. 

The most interesting feature of Fig.~\ref{Loss_rate} is the dependence of the cooling efficiency on the rate at which the truncated energy changes. It is known that the evaporation removes energetic atoms from the trap and this has a cooling effect. In contrast the three-body recombination loss has a heating effect, since it selectively removes atoms with low energy from the highest density part of the system. The cooling efficiency is determined by the ratio between these two effects. With a fast RF sweep, the loss rate caused by evaporation (A1) is larger than the loss rate caused by three-body recombination (A2) throughout most of the evaporative cooling process. Both the large cooling efficiency and the quick production of the condensate enable the evaporative cooling to be highly efficient and the total $N_{0}$ reduction to be small. In contrast, with a medium (slow) RF sweep, the loss rate caused by three-body recombination, B2 (C2), exceeds the loss rate caused by evaporation, B1 (C1), and it takes a long time for condensate growth to finish. For this reason the medium (slow) RF sweep provides inefficient evaporative cooling and the total $N_{0}$ reduction is larger than that with a fast RF sweep. 

\section{Conclusion}
%
We have succeeded in demonstrating the efficient rapid production of a large Bose-Einstein condensate in $^{87}$Rb atoms with a fast linear RF sweep during the final stage of evaporative cooling against serious three-body loss. By measuring the growth of a condensate, we were able to study the dependence of the cooling efficiency on the rate at which the truncated energy changes. The experimental result quantitatively agrees with a calculation based on the quantum kinetic theory of a Bose gas.

In this paper, only the RF sweeps were optimized to improve the cooling efficiency. It is also known that the confinement of magnetic trap is an important parameter to control the serious three-body loss rate by changing the atomic gas density. A promising future work for obtaining a further increase in $N_{0}$ in $^{87}$Rb atoms is the optimization of both the RF sweep and the trapping confinement at the same time.

\section*{Acknowledgments}
%
The authors thank T. Hong of the University of Washington, M. Ueda of Tokyo Institute of Technology, F. Shimizu of Institute for Laser Science, M. W. Jack of Rice University, M. Koashi of Osaka University, N. Imoto of the Graduate University for Advanced Studies, M. Mitsunaga of Kumamoto University, Takaaki Mukai of NTT Electronics Corporation, and Y. Tokura of NTT Basic Research Laboratories for valuable discussions.

%
\newcounter{q}
\setcounter{q}{118}

\newpage 
\bibliography{apssamp}

\end{document}